\documentclass[twocolumn,showpacs,amsmath,amssymb,pra]{revtex4}

\usepackage{epsfig}

\def\rmd{{\rm d}}
\newcommand{\sgn}{\mathop{\rm sgn}}
\newcommand{\pFu}{p_{Fh}}
\newcommand{\pFd}{p_{Fl}}
\newcommand{\mup}{m_h}
\newcommand{\mdwn}{m_l}
\newcommand{\rhou}{\rho_h}
\newcommand{\rhod}{\rho_l}
\newcommand{\Eu}{E_h}
\newcommand{\Ed}{E_l}
\newcommand{\eref}[1]{(\ref{#1})}

\begin{document}
\title{Superfluidity in the interior-gap states}
\author{Shin-Tza Wu and Sungkit Yip}
\affiliation{
Institute of Physics, Academia Sinica, Nankang, Taipei 115,
Taiwan}
\date{March 10, 2003}

\begin{abstract}
We investigate superfluidity in the interior-gap states proposed
by Liu and Wilczek. At weak coupling, we find the {\em gapless}
interior-gap state unstable in physically accessible regimes of the
parameter space, where the superfluid density is shown to be always
negative. We therefore conclude that the spatially-uniform interior-gap
phase is extremely unstable unless it is fully gapped; in this case,
however, the state is rather similar to conventional BCS states.
\end{abstract}
\maketitle

\section{Introduction}
The advances of techniques in manipulating dilute cold atoms in traps
have opened up many fascinating new possibilities to condensed matter
physics in recent years.\cite{BEC} The Bose-Einstein condensation of dilute
alkali atoms has provided new Bose-condensed systems previously known
only for liquid $^4$He. Experimentalists are now investigating the
possibility of realizing the Bardeen-Cooper-Schrieffer (BCS) states in
trapped Fermionic atoms.\cite{fermion1,fermion2,fermion3} In the simplest
case of
spin-singlet BCS states, as in ordinary superconductors such as aluminum,
the ``conventional" Cooper pairs are formed from Fermions with
opposite spins.\cite{Tinkham} In atom traps, a corresponding realization
could be two species of Fermions which pair via the inter-species
interaction. The two species of Fermions can be the same kind of atoms
(say, $^6$Li) in different hyperfine states, or, in the most general case,
different Fermionic atoms.

Conventionally, the BCS state is a condensate of Cooper pairs
consisting of Fermions with equal mass and opposite spins from states
with a single Fermi surface. In considering superconductivity in
systems containing ferromagnetically coupled paramagnetic impurities,
Fulde and Ferrell, and independently Larkin and Ovchinnikov\cite{FFLO}
(FFLO), studied Cooper-pairing in systems with mismatched Fermi
surfaces. They found a spatially {\em non-uniform} phase can be more
favored than the BCS states in certain ranges of temperature and
Fermi-surface mismatches. Recently Liu and Wilczek (LW), motivated by
recent developments in atomic physics, studied a system of interacting
Fermions consisting of two species of particles with unequal masses
and mismatched Fermi surfaces.\cite{LW} Assuming pairing only in the
vicinity of the smaller Fermi surface, they claimed that at weak
coupling there exists a novel superconducting state, which is
spatially uniform, could be energetically more favorable than the FFLO
state if the coupling strength is above a critical value.  This state
is characterized by the coexistence of superfluid and normal
components {\em at zero temperature} that are separated by a momentum
gap between the two Fermi surfaces (thus the ``interior-gap"
state). LW concluded that the critical coupling strength for the
interior-gap state could be vanishingly small when the mass difference
between the two Fermion species is large.

In this paper, we re-examine the possibility of an interior-gap phase
having in mind cold dilute Fermionic atoms in traps. Instead of
postulating pairing only around the smaller Fermi surfaces,\cite{LW} here
we consider a short-range attractive interaction arising from the
low energy scattering between Fermions of unlike species. We derive and
solve the gap equation, and then further examine the current response of
the interior-gap state in the weak coupling limit. Surprisingly, we find
the state unstable in all physically accessible regimes unless it is fully
gapped.

\section{The model and the gap equation}
We consider an atom trap which contains two species of Fermions with
mass $m_h>m_l$. We assume that the gas is dilute, with the Fermions
interacting via a short-range interaction ({\it i.e.}, both the scattering
length and the range of the interaction between the two Fermion species
are much less than the interparticle distances). This interaction can
therefore be modeled by a delta-function like coupling of strength $-g>0$
which ultimately causes pairing between the heavy ($h$) and the light ($l$)
particles. The system is thus described by the Hamiltonian
\begin{eqnarray}
H = \sum_{{\bf p}\, \alpha=h,l} \xi_{{\bf p}\alpha}
a_{{\bf p}\alpha}^\dag a_{{\bf p}\alpha}
+\frac{g}{2}\sum_{{\bf p},{\bf p}'}
a_{{\bf p}'h}^\dag a_{-{\bf p}'l}^\dag
a_{-{\bf p}l} a_{{\bf p}h} .
\end{eqnarray}
Here $a_{{\bf p}\alpha}$ and $a_{{\bf p}\alpha}^\dag$ annihilates
and creates an $\alpha$ species particle with momentum ${\bf p}$. For
simplicity, we ignore the trapping potential and take the parabolic
dispersions for the particles
$\xi_{{\bf p}\alpha}=(p^2-p_{F\alpha}^2)/2m_\alpha$ with $p_{F\alpha}$
the Fermi momentum.

Since the gas is dilute, we shall make mean-field approximation and define the
space-independent order parameter
$\Delta=-g\sum_{\bf p}\langle a_{{\bf p}h}a_{-{\bf p}l}\rangle$
which is chosen to be real. The quasiparticle dispersions are then obtained
from the standard procedure,\cite{Tinkham} yielding
\begin{eqnarray} \label{Ek}
E_{{\bf p}h,l}
= \pm\frac{\xi_{{\bf p}h}-\xi_{{\bf p}l}}{2}
+ \sqrt{\left(\frac{\xi_{{\bf p}h}+\xi_{{\bf p}l}}{2}\right)^2
+ \Delta^2} \, .
\end{eqnarray}
The gap equation is derived from the Hamiltonian upon minimizing the
free energy with respect to $\Delta$. However, due to the short-range
nature of the pairing interaction, the resulting formula has ultraviolet
divergence. To resolve the problem, we eliminate the coupling constant
$g$ in favor of the scattering length $a$ for two-particle
 scattering (between $h$ and $l$) in vacuum.
The regularized gap equation reads\cite{randeria}
\begin{eqnarray} \label{gap_eqn}
-\frac{m_r}{2\pi a} \!=\!\sum_{\bf p} \!\!
\left[
\frac{1}{\Eu+\Ed}(1-f(\Eu)-f(\Ed))
-\frac{1}{\xi_h^0+\xi_l^0}
\right] ,
\end{eqnarray}
where $m_r=\mup\mdwn/(\mup+\mdwn)$ is the reduced mass, $a$ the $s$-wave
scattering length; $f$ is the Fermi function and
$\xi_\alpha^0=p^2/2m_\alpha$ the particle dispersions in vacuum. Here
and in below, for brevity, we omit the subscript ${\bf p}$ when no
confusion would likely arise.
\begin{figure}
\hspace*{-10mm}
\includegraphics*[width=90mm]{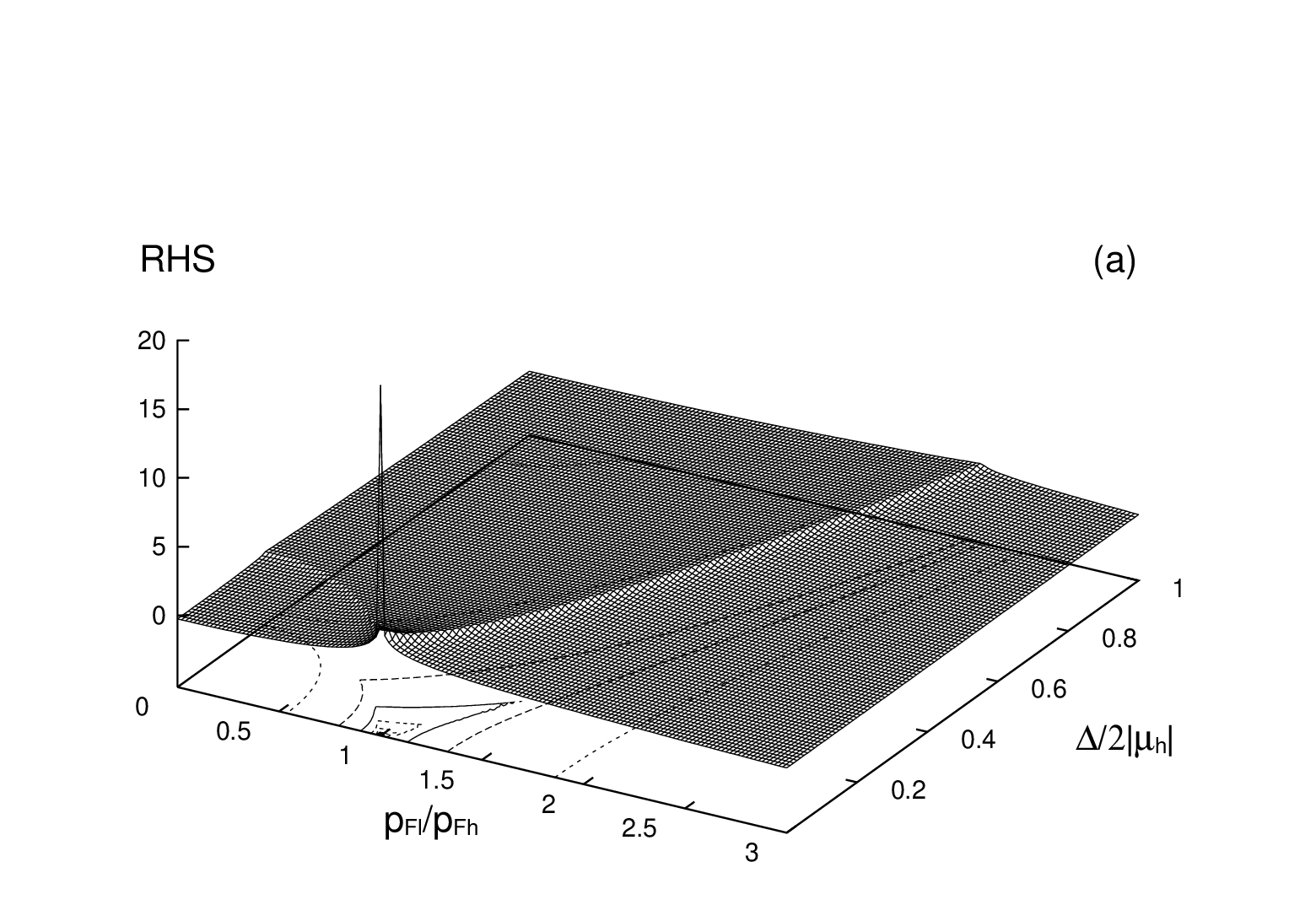}\\
\includegraphics*[width=80mm]{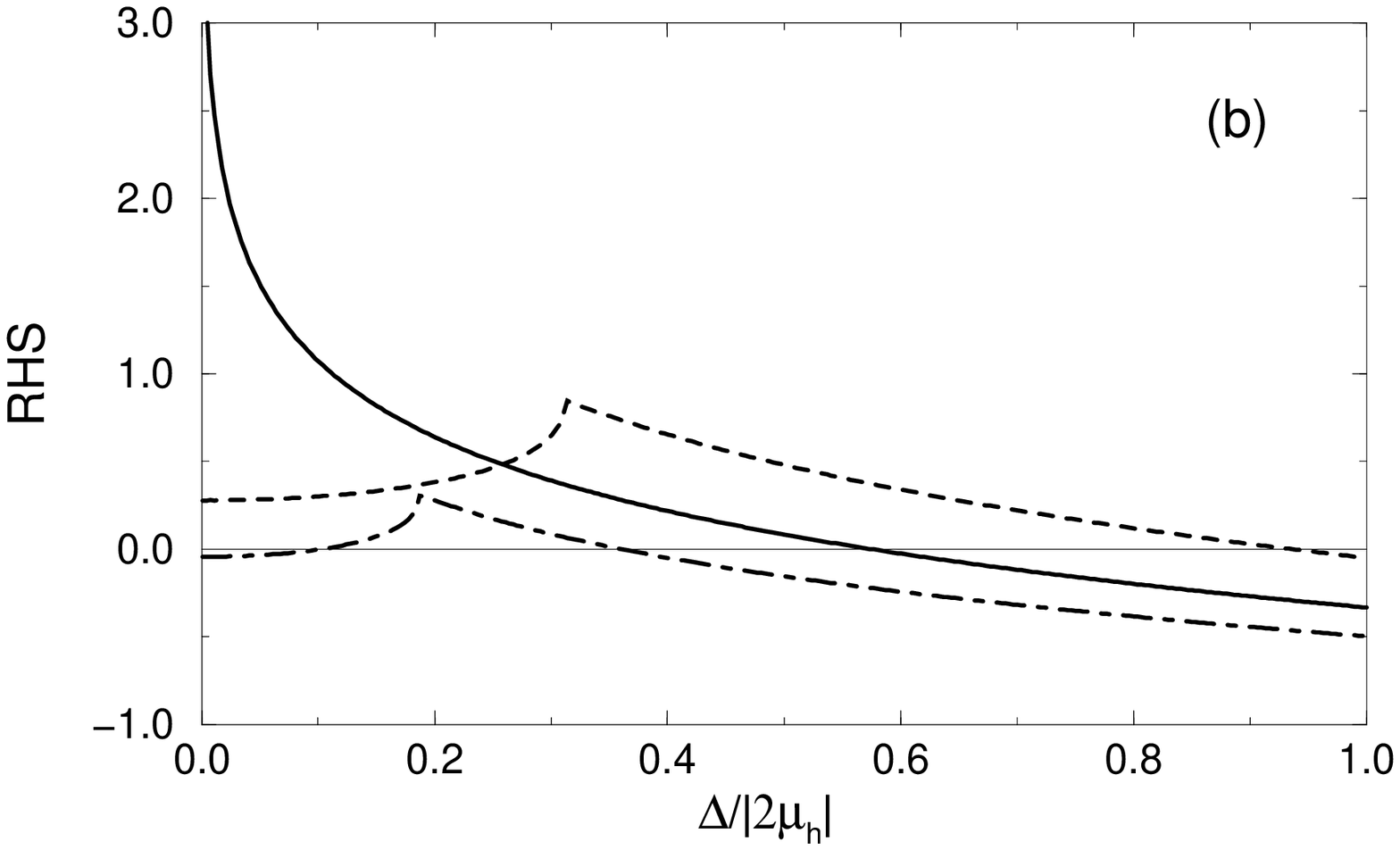}
\caption{\small The right hand side of the gap equation
[in units of $2\pi/(p_{Fh}m_r$)] for
$\mup=\mdwn$; $(a)$ shows the three-dimensional plot, and $(b)$ the
profiles at $\pFd/\pFu$=0.5 (dot-dashed line), 1.0 (full line), and
1.5 (dashed line). The ridges in $(a)$ correspond to the critical value
$\Delta_c$
in Eq.~\eref{Dc} which appear as peaks in $(b)$.}
\label{RHS_plot}
\end{figure}
\begin{figure}
\hspace*{-10mm}
\includegraphics*[width=80mm]{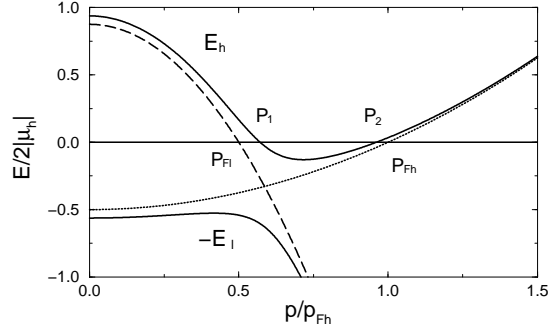}\\
\caption{\small The quasiparticle dispersions $E_h$ and $-E_l$ for
$\mup=7\mdwn$ and $\pFu=2.0\pFd$; the normal-state dispersions $\xi_h$
and $-\xi_l$ are also shown for the heavy (dotted line) and the light
(dashed line) species. Here $\Delta=0.6|\mu_h| < \Delta_c = 0.99 |\mu_h|$.
It is seen that $E_h$ intersects the $p$-axis at the points $p_1$ and
$p_2$, which correspond to the regions of gapless excitations.}
\label{dispersion}
\end{figure}

We solve the gap equation at zero temperature by plotting the right hand
side (RHS) of Eq.~\eref{gap_eqn} for different values of $\mdwn/\mup$
over the parameter space $(\bar{p}_F,\bar{\Delta})$, where
$\bar{p}_F\equiv\pFd/\pFu$ and $\bar{\Delta}\equiv|\Delta|/2|\mu_h|$ with
$\mu_h$ the chemical potential of the heavy particles. Since $a<0$ for
attractive interactions (without formation of $h$-$l$ bound states in
vacuum), we seek solutions of $\Delta$ such that the
RHS is positive. Figure \ref{RHS_plot} shows typical plots for the RHS of
the gap equation with $\mup=\mdwn$. The profile at $\pFd/\pFu$=1 in
Fig.~\ref{RHS_plot}$(b)$ thus corresponds to the usual BCS case. In this
case the RHS diverges as $\Delta\to 0$, corresponding to the well-known
fact that an infinitesimally small attractive interaction can lead to
pairing. However, two differences from the usual BCS states occur here.
Firstly, the RHS becomes negative at large $\bar{\Delta}$, indicating no
solution for attractive interaction. Secondly, when $\pFu\neq\pFd$, as
displayed in Fig.~\ref{RHS_plot}$(b)$, a finite peak (maximum as a
function of $\Delta$) arises in the RHS of the gap equation at finite
$\Delta$.
In the three-dimensional plots of Fig.~\ref{RHS_plot}$(a)$, these peaks
appear as ridges over the $(\bar{p}_F,\bar{\Delta})$ plane.
The existence of these peaks (ridges) implies that the system requires a
critical coupling strength for solutions to the gap equation when there
are Fermi surface mismatches. Another significance of these peaks and
ridges is related to the quasiparticle properties of the system, as we
shall now explain.

An important feature of the interior-gap state as proposed by LW is the
coexistence of the superfluid and the normal components
{\em at zero temperature}; in other words, there exists gapless
excitations. However, as one can check from Eq.~\eref{Ek}, this is
possible only when the magnitude of the order parameter $|\Delta|$
is smaller than
\begin{equation}\label{Dc}
\Delta_c =
\frac{|p_{Fh}^2-p_{Fl}^2|}{4\sqrt{\mup\mdwn}} \, .
\end{equation}
When $|\Delta| < \Delta_c$, depending on the relative magnitude of
$\pFu$ and $\pFd$, either $\Eu$ or $\Ed$ crosses zero at the points
(see Fig.~\ref{dispersion})
\begin{equation}\label{p1p2}
p_{1,2} = \left(
\frac{\pFu^2+\pFd^2\mp\sqrt{(\pFu^2-\pFd^2)^2-16\mup\mdwn\Delta^2}}{2}
\right)^{1/2} .
\end{equation}
In the event $|\Delta|>\Delta_c$, $E_{h,l}$ both stay positive for all
values of $p$. There are then no gapless excitations and the
interior-gap state has only the superfluid component at zero temperature.
The peaks and ridges in the plots for the RHS turn out to locate
exactly at the value $|\Delta|=\Delta_c$ of Eq.~\eref{Dc}. Indeed since
$|\Delta|=\Delta_c$ draws the boundary between the gapped and the gapless
regions over the parameter space, one can expect a qualitative change
when plotting the RHS across this line.
 We will be interested mainly in the gapless regions, namely where
$|\Delta|<\Delta_c$.

\section{Current response}
When $|\Delta|<\Delta_c$ the interior-gap state has gapless excitations
from the quasiparticle branch whose dispersion crosses zero.
In the FFLO ground state, the quasiparticles produce a flow which cancels
exactly the current due to the finite-momentum Cooper pairs. This lowers
the free energy and stabilizes the FFLO states.\cite{FFLO} In the
interior-gap state, the Cooper pairs are stationary in the absence of
any superflow. It is therefore of interest to
investigate the current response of the interior gap state; in particular,
the effects of the quasiparticles in the current-carrying states.

In the static case the quasiparticle distributions of the interior-gap
phase are given by the expression
\begin{eqnarray} \label{n_eqlbm}
n_{{\bf p}\alpha}^0 =
u_{\bf p}^2 f(E_{{\bf p}\alpha}) +
v_{\bf p}^2 f(-E_{{\bf p}\bar{\alpha}}) \, ,
\end{eqnarray}
where $\bar{\alpha}$ is the species other than $\alpha$ and
the coherence factors are
\begin{eqnarray} \label{uk_vk}
u_{\bf p}^2=\frac{\Eu+\xi_l}{\Eu+\Ed}=\frac{\Ed+\xi_h}{\Eu+\Ed}\, ,
\quad
v_{\bf p}^2=1-u_{\bf p}^2 \, .
\end{eqnarray}
In the presence of a small superfluid velocity ${\bf w}$, the
quasiparticle energies are shifted by $+({\bf p \cdot w})$ so that
the quasiparticle distribution functions of Eq.~\eref{n_eqlbm} become
\begin{eqnarray} \label{n_pw}
n_{{\bf p}\alpha} =
u_{\bf p}^2 f(E_{{\bf p}\alpha}+{\bf p \cdot w}) +
v_{\bf p}^2 f(-E_{{\bf p}\bar{\alpha}}+{\bf p \cdot w}) \, .
\end{eqnarray}
The number current can be decomposed as usual into two parts\cite{Tinkham}
\begin{eqnarray}
{\bf J}_\alpha^p &=&
\frac{1}{m_\alpha} \sum_{\bf p}
n_{{\bf p}\alpha} {\bf p}
\equiv \rho_\alpha^p  {\bf w} \, ,
\label{p_flow}
\\
\label{d_flow}
{\bf J}_\alpha^d &=& \sum_{\bf p}
n_{{\bf p}\alpha} {\bf w}
\equiv \rho_\alpha^d  {\bf w} \, ,
\end{eqnarray}
where the superscripts $p$ and $d$ indicate the paramagnetic and
the diamagnetic components; $\rho_\alpha^{p,d}$ are the corresponding number
densities. In the following we shall examine Eqs.~\eref{p_flow} and
\eref{d_flow} in the limit of small ${\bf w}$ at zero temperature.

Let us consider the case when $\pFu>\pFd$ and $|\Delta|<\Delta_c$.
Then, as can be checked from \eref{Ek}, $\Eu<0$ for $p_1<p<p_2$ while
$\Ed>0$ always. Therefore the $l$ branch of quasiparticles is
always empty, {\it i.e.}, $f(E_l)=0$ and $f(-E_l)=1$, at zero temperature.
In the presence of a small superfluid
velocity $\bf w$, since $E_l$ never changes sign, one has
\begin{eqnarray} \label{f_El}
f(\pm E_l+{\bf p}\cdot{\bf w}) - f(\pm E_l)  = 0 .
\end{eqnarray}
Using
the fact that ${\bf J}_\alpha^p=0$ when ${\bf w}=0$, we obtain for small
superfluid velocity the paramagnetic currents
\begin{eqnarray}
{\bf J}_\alpha^p
&=& \frac{1}{m_\alpha} \sum_{\bf p} {\bf p}
(n_{{\bf p}\alpha}-n_{{\bf p}\alpha}^0)
\nonumber \\
&\simeq&\frac{1}{m_\alpha}\!\sum_{\bf p}
{\bf p} \, \phi_\alpha^2 \left[
f(\eta_\alpha E_h+{\bf p}\cdot{\bf w}) - f(\eta_\alpha E_h)
\right] \!,
\label{Jp}
\end{eqnarray}
where for the heavy and the light branches
\begin{eqnarray}
\phi_\alpha =
   \left\{
          \begin{array}{l}
          u_{\bf p} \\ v_{\bf p}
          \end{array}
   \right.
\quad {\rm and} \quad
\eta_\alpha =
   \left\{
          \begin{array}{l}
          +1 \\ -1
          \end{array}
   \right.
\quad \mbox{\rm for $\alpha=$}
   \left\{
          \begin{array}{c}
          h. \\ l.
          \end{array}
   \right.
\end{eqnarray}
In arriving at the final expression in Eq.~\eref{Jp}, we have used
\eref{f_El} for small $\bf w$.

To leading order in the superfluid velocity, the term in the square
brackets in Eq.~\eref{Jp} is proportional to the delta function
$\delta(\Eu)$, which vanishes except at $p=p_1$ and $p=p_2$. Replacing the
sum by integral, one thus obtains
\begin{eqnarray} \label{rho_p}
\rho_\alpha^p = - \frac{1}{6\pi^2}\frac{1}{m_\alpha}
\left(p_1^4 D_1 \phi_{\alpha 1}^2
+ p_2^4 D_2 \phi_{\alpha 2}^2 \right) \, ,
\end{eqnarray}
where $D_i\equiv 1/|\partial \Eu/\partial p|_i$ and the subscripts $i=1,2$
indicate evaluating at the points $p_1$, $p_2$. Note that
$\rho_\alpha^p$ is always negative, which means that the paramagnetic
current ${\bf J}_\alpha^p$ always flows in the opposite direction to
${\bf w}$.

To leading order in ${\bf w}$, the diamagnetic current is simply
\begin{eqnarray}
{\bf J}_\alpha^d=\sum_{\bf p}n_{{\bf p}\alpha}^0{\bf w} \, ,
\end{eqnarray}
which is always in the same direction as ${\bf w}$. When $\pFu>\pFd$ the
diamagnetic number-densities can be expressed as
\begin{eqnarray} \label{rho_d}
\rho_{h,l}^d
= \frac{p_{2,1}^3}{6\pi^2}+
\frac{1}{2\pi^2}
\left(-\int_0^{p_1}\!\!\!\rmd p \, p^2 u^2 +
\int_{p_2}^\infty \!\!\!\rmd p \, p^2 v^2 \right) .
\end{eqnarray}
Note that $\rho_\alpha^d$ is always positive since the diamagnetic
current ${\bf J}_\alpha^d$ is always in the same direction as ${\bf w}$.

Similar analysis as above can also be done for the case of $\pFu<\pFd$.
As it turns out, we find that the total superfluid number densities
$\rho_\alpha^n\equiv\rho_\alpha^p+\rho_\alpha^d$ for the two species are
always identical (see Appendix). In other words, despite
the seemingly asymmetric properties of the two species, the
superflow turns out to be quite conventional -- the particles flow together
in pairs. The superfluid mass density is thus simply
$\rho_M\equiv\mup\rhou^n+\mdwn\rhod^n=(\mup+\mdwn)\rhou^n$.

\begin{figure}
\includegraphics*[width=80mm]{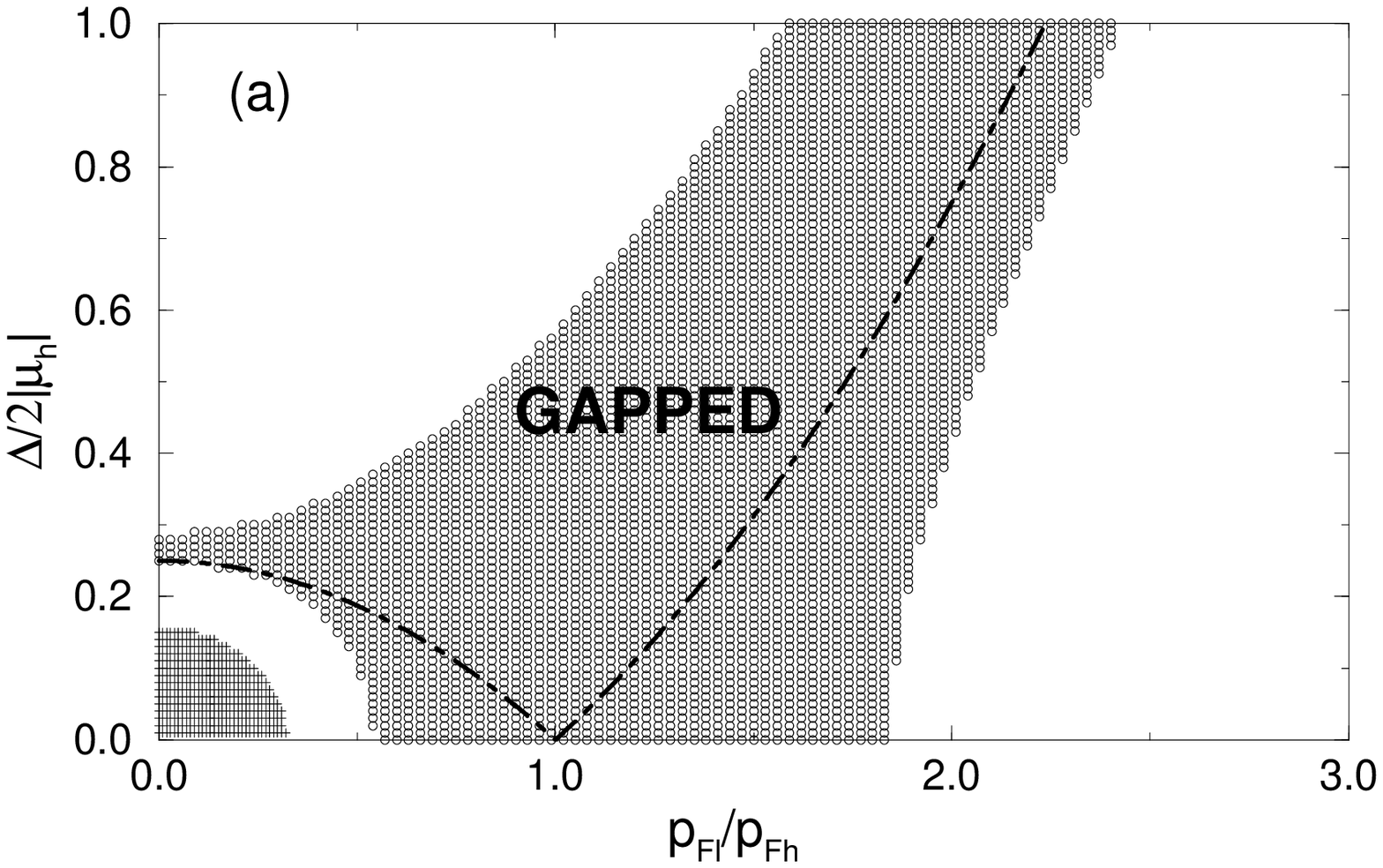} \\
\includegraphics*[width=80mm]{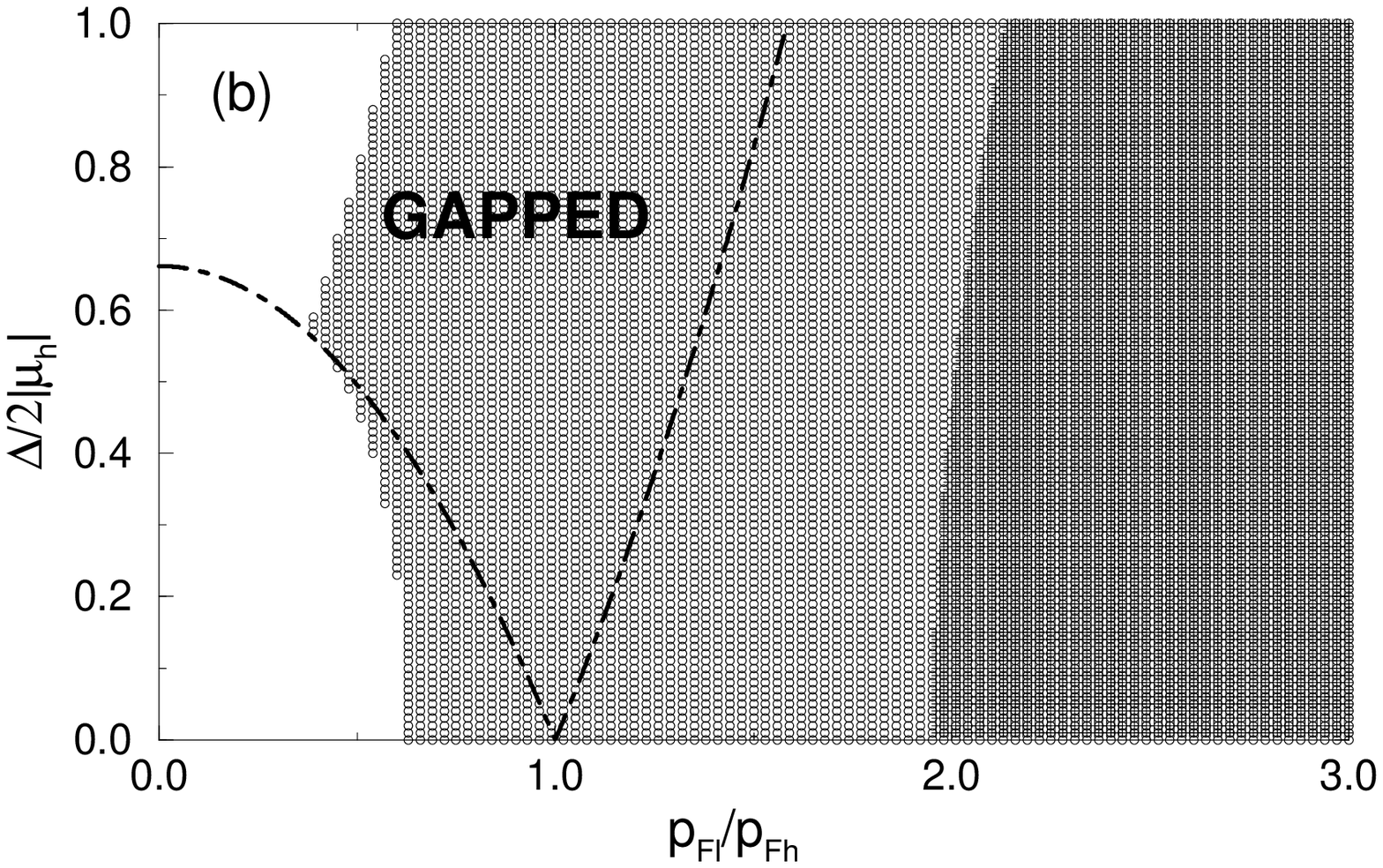} \\
\includegraphics*[width=80mm]{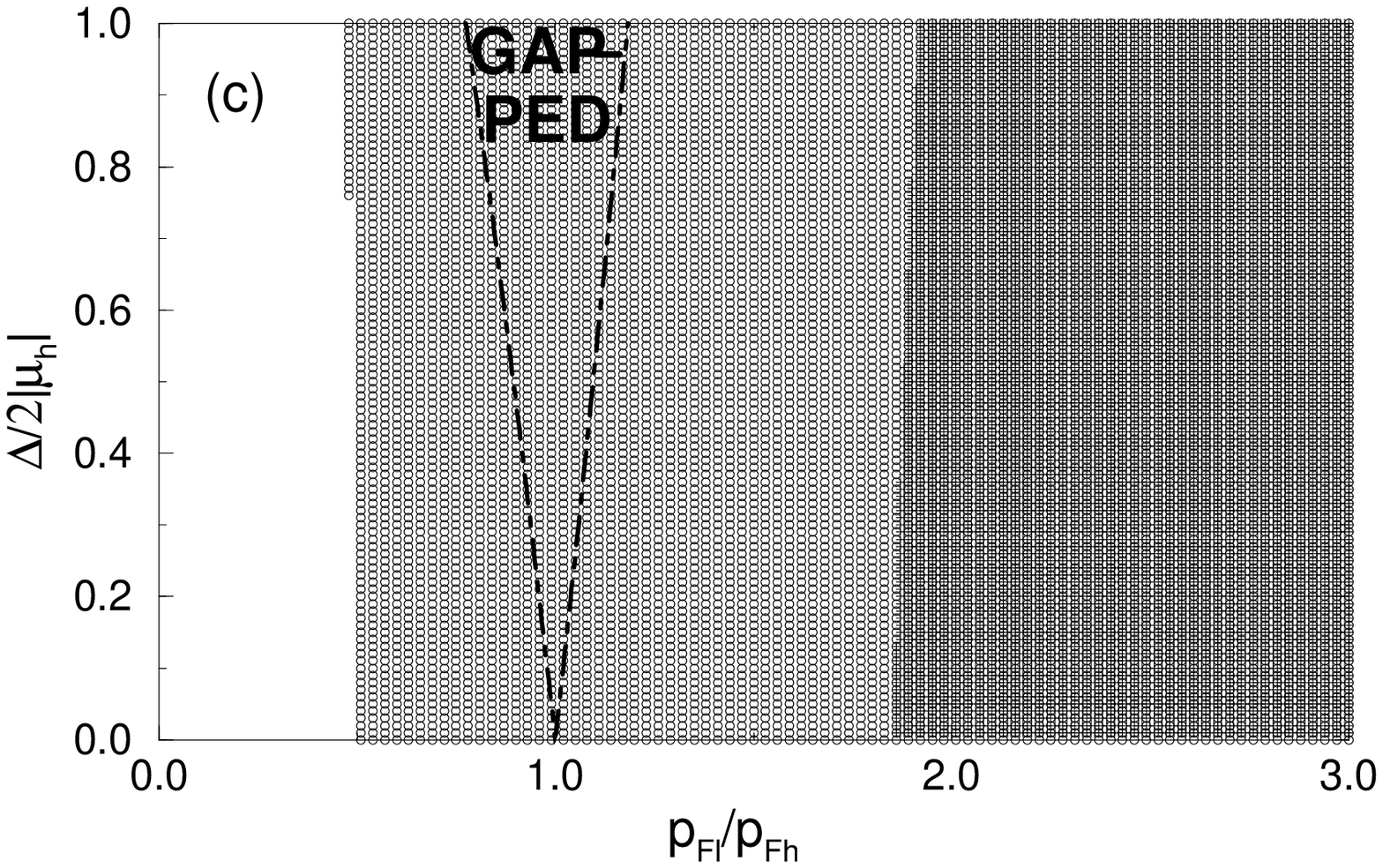} \\
\caption{\small Summary of our results for the interior-gap states for
$\mdwn/\mup=$ $(a)$ 1, $(b)$ 1/7, and $(c)$ 1/100 over the 
$(\bar{p}_F,\bar{\Delta})$ plane. The grey areas are regions where
the gap equation has solutions for attractive interactions. The dark areas
mark the regions where the superfluid mass density is positive {\em in the
gapless regions}. The dash-dotted line depicts the critical line determined
from Eq.~\eref{Dc}, which also corresponds to the ridges in the RHS of the
gap equation (see Fig.~\ref{RHS_plot}$(a)$).}
\label{phase}
\end{figure}
\begin{figure}
\includegraphics*[width=80mm]{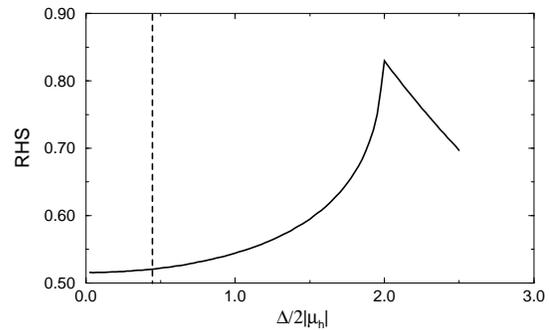}
\caption{\small The RHS of the gap equation [in units of $2\pi/(p_{Fl}m_r$)]
for $\mup=7\mdwn$ at $\pFd=2\pFu$. The
dashed line indicates the value of $\bar{\Delta}$ below which the superfluid
mass density is positive (cf.~Fig.~\ref{phase}$(b)$). To the right of the
peak, the state becomes fully gapped, so that the mass density is
again positive trivially. }
\label{RHS_profile}
\end{figure}

As noted earlier, $\rho_\alpha^d$ is always positive while
$\rho_\alpha^p$ is always negative; as a result, the sign of $\rho_M$
is determined from the competition between the paramagnetic and the
diamagnetic components. Quite unexpectedly, we find the superfluid
mass density of the interior-gap phase stays negative in almost all
regions of the parameter space. This can also be confirmed
analytically for small $\Delta$. A negative superfluid mass density
implies that the interior-gap phase is unstable towards a
spontaneously generated superfluid velocity or a phase gradient, since
the kinetic energy is $\frac{1}{2}\rho_M{\bf w}^2<0$.  Our results for
$\mdwn/\mup=$1, 1/7, and 1/100 are shown in Fig.~\ref{phase}. From
Fig.~\ref{phase}$(a)$ one observes that for $\mup=\mdwn$ the regions
where $\rho_M>0$ and the regions where the gap equation has solutions
are completely disjoint to each other in the gapless region. Namely
for $\mup=\mdwn$, except in the fully gapped region, there is no where
in the parameter space $(\bar{p}_F,\bar{\Delta})$ the interior-gap
state being stable. This is consistent with the FFLO
results.\cite{FFLO} For larger mass ratio at $\mup=7\mdwn$, as shown
in Fig.~\ref{phase}$(b)$, there are regions in the gapless area where
$\rho_M$ stays positive and the gap equation has solutions. To
estimate the coupling strength required, we plot in
Fig.~\ref{RHS_profile} the profile of the RHS of the gap equation at
$\pFd=2\pFu$. Disregarding the fact that there are two solutions to
the gap equation in this case (see next Section for a discussion), we
find the coupling strength necessary for $\rho_M>0$ around $|\pFd
a|\sim 1.9$. At such strong couplings, however, the validity of the
present mean-field calculation is doubtful. For even larger mass ratio
at $\mup=100\mdwn$, the situation stays much the same (see
Fig.~\ref{phase}$(c)$); at $\pFd=2\pFu$ we find $|\pFd a|\sim 0.7$ for
$\rho_M$ to be positive. Even though the coupling strength does go
down with increasing mass ratio, it is unrealistic to consider even
higher values of $\mup/\mdwn$. Note also that the regions of $\rho_M >
0$ occurs for $p_{Fl} > p_{Fh}$, instead of $p_{Fl} < p_{Fh}$ as
suggested by LW.\cite{LW}

From the sign of the superfluid density, our results show that in the
interior-gap phase the paramagnetic component overwhelms the
diamagnetic component whenever there are gapless excitations. It is
only when the quasiparticles are gapped out the number density in the
interior-gap state would stay positive at weak couplings.

\section{Summary and discussion}
Within the range of validity of our calculation, we find the interior-gap
phase unstable in physically accessible regions of the parameter space;
the superfluid mass density is always negative at weak couplings,
rendering a uniform phase unstable. Therefore, starting
from the normal state, upon increasing coupling strength the system
considered here would/may first enter an inhomogeneous FFLO state which
persists until the onset of a fully gapped ``interior-gap" state;
a uniform {\em gapless} interior-gap state has no room in the intervening
regimes. Indeed this may have been expected from continuity: As was noticed,
due to the existence of the peaks and ridges in the RHS, there can be two
solutions to the gap equation (see Fig.~\ref{RHS_plot}). Since at
$\pFu=\pFd$ the only solution is always in the gapped region, from
continuity one expects the stable solutions when $\pFu\neq\pFd$ to be
those above the V shape boundaries in Fig.~\ref{phase}. In other words,
continuity implies that it is always the gapful solutions that are stable.
This includes the large $|p_{Fl} a|$ regimes discussed near the end of the
last section. For this fully gapped phase, there are no quasiparticles
at zero temperature and the number densities of the $h$ and the $l$
branches are always equal.\cite{note2} This state thus behaves just like
an ordinary BCS state.

This research was supported by NSC of Taiwan under grant
number NSC 91-2112-M-001-063.

\vspace*{5mm}
\section*{Appendix}
\setcounter{section}{1}
\setcounter{equation}{0}
\renewcommand{\theequation}{\Alph{section}$\,$\arabic{equation}}
In this appendix we show that the superfluid number densities for the
heavy and the light species are identical at small superfluid
velocities.

Let us consider first the case of $\pFu>\pFd$. We note first that from
Eqs.~\eref{Ek} and \eref{uk_vk}
\begin{eqnarray} \label{dEdp}
\frac{\partial \Eu}{\partial p}
= p \left(\frac{u^2}{\mup} - \frac{v^2}{\mdwn} \right) \, .
\end{eqnarray}
Substituting the above expression into Eq.~\eref{rho_p},
we obtain the difference between the paramagnetic number densities of
the two species\cite{note1}
\begin{eqnarray} \label{rhop_diff1}
\rhou^p-\rhod^p = -\frac{1}{6\pi^2}
&&\left[
p_1^3  \sgn\left(\frac{u_1^2}{\mup}-\frac{v_1^2}{\mdwn}\right)
\right.\nonumber\\&&\left.
+p_2^3 \sgn\left(\frac{u_2^2}{\mup}-\frac{v_2^2}{\mdwn}\right)
\right] \,.
\end{eqnarray}
Noting that $\Eu(p_{1,2})=0$, one can derive from \eref{uk_vk}
\begin{eqnarray}
\frac{u_1^2}{v_1^2}<\frac{\mup}{\mdwn} \, ,
\qquad
\frac{u_2^2}{v_2^2}>\frac{\mup}{\mdwn} \, .
\end{eqnarray}
Therefore \eref{rhop_diff1} becomes
\begin{eqnarray}\label{rhop_diff}
\rhou^p-\rhod^p = - \frac{1}{6\pi^2}
\left( p_2^3-p_1^3 \right) \, .
\end{eqnarray}

For the diamagnetic number densities, at small superfluid
velocity the difference between the two species is simply that of their
quasiparticle occupation numbers. As one can find trivially from
Eq.~\eref{rho_d}
\begin{eqnarray} \label{rhod_diff}
\rhou^d-\rhod^d = + \frac{1}{6\pi^2}
\left( p_2^3-p_1^3 \right) \,.
\end{eqnarray}
Combining Eqs.~\eref{rhop_diff} and \eref{rhod_diff}, we conclude that
$\rhou^n-\rhod^n=0$, namely that the superfluid number densities of the two
species are identical.

For $\pFu<\pFd$ similar calculations as above lead to
\begin{eqnarray}
\rhou^p-\rhod^p &=& + \frac{1}{6\pi^2}
\left( p_2^3-p_1^3 \right) \, ,
\\
\rhou^d-\rhod^d &=& - \frac{1}{6\pi^2}
\left( p_2^3-p_1^3 \right) \,.
\end{eqnarray}
Therefore, again, one finds $\rhou^n=\rhod^n$ in this case.


\begin{thebibliography}{99}
\bibitem{note1}
One may notice that when $u_i^2/\mup=v_i^2/\mdwn$, Eq.~\eref{dEdp}
vanishes and the expression \eref{rho_p} becomes singular. However, this
takes place only when the dispersion curve is tangent to the energy zero,
{\it i.e.}~when $|\Delta|=\Delta_c$. Since we are interested in the case
of $|\Delta|<\Delta_c$, this shall not concern us.
\bibitem{note2}
This means that there must be phase separation in this fully gapped phase,
since in general the $h$ and the $l$ particles can have unequal numbers in
the normal state.
\end{thebibliography}
\end{document}